\documentclass[reprint,showpacs,preprintnumbers,amsmath,amssymb,floatfix]{revtex4-1}
\usepackage{float}
\usepackage{color}
\usepackage{graphicx}
\usepackage{dcolumn}
\usepackage{bm}
\usepackage{xcolor}
\usepackage{comment}
\usepackage{multirow}
\usepackage[colorlinks,citecolor=blue,linkcolor=red,anchorcolor=blue,filecolor=blue,urlcolor=blue]{hyperref}
\usepackage{color}
\usepackage{csquotes}
\usepackage{comment}
\begin{document}
\title{Nuclear incompressibility and its enduring impact on fusion cross-section}
\author{Shilpa Rana$^1$}
\email{srana60\_phd19@thapar.edu}
\author{M. Bhuyan$^2$}
\email{bunuphy@um.edu.my}
\author{S. K. Patra$^{3,4}$}
\author{Raj Kumar$^1$}
\email{rajkumar@thapar.edu}
%
\affiliation{$^1$Department of Physics and Materials Science, Thapar Institute of Engineering and Technology, Patiala, Punjab 147004, India}
\affiliation{$^2$Center for Theoretical and Computational Physics, Department of Physics, \\ Faculty of Science, Universiti Malaya, Kuala Lumpur 50603, Malaysia}
\affiliation{$^3$Institute of Physics, Sachivalya Marg, Bhubaneswar-751005, India}
\affiliation{$^4$Homi Bhabha National Institute, Training School Complex, Anushakti Nagar, Mumbai 400094, India}
\bigskip 
\begin{abstract}
The fusion mechanism of reactions involving even-even $^{112-124}$Sn, doubly magic $^{132}$Sn, $^{208}$Pb as targets, and $^{64}$Ni as the projectile is explored within the relativistic mean field (RMF) formalism. The main aim of choosing these nuclei is to explore the correlation between the nuclear incompressibility and the fusion cross-section. The nucleus-nucleus interaction potential is calculated by folding the axially deformed nuclear densities and the relativistic R3Y nucleon-nucleon (NN) potential obtained for the nonlinear sets of NL3$^*$, hybrid, and NL1, which yield different values for various characteristics of nuclear matter at saturation. The fusion barrier characteristics obtained for different RMF parameterizations are further used to calculate the cross-section within the $\ell$-summed Wong model. We found a decrease in the barrier height and consequently, an increase in the cross-section with a decrease in the incompressibility for all sets of parameters considered. Furthermore, comparing the barrier heights obtained for NL3$^*$ and the hybrid parameters, it is observed that the barrier height decreases with decreasing symmetric energy and incompressibility value. Moreover, a lower barrier height and, consequently, a higher cross-section at below-barrier energies is observed for the NL1 parameter set, which gives a soft equation of state (EoS) having a lower value of nuclear matter incompressibility. The calculated cross section is satisfactorily consistent with the available experimental data for $^{64}$Ni+$^{208}$Pb system. In contrast, the nuclear potentials obtained for NL3$^*$ and the hybrid parameter sets underestimate the cross-section at below-barrier energies for $^{64}$Ni+$^{112-124,132}$Sn reactions. This discrepancy between the experimental data and the theoretical results for $^{64}$Ni+$^{112-124,132}$Sn reactions can be correlated with the soft behaviour of the Sn isotopes. The compressible nature of Sn-isotopes is inferred to lower the barrier height, which further leads to enhancement of the experimental fusion and/or capture cross-section at below-barrier energies. Thus, the NL1 parameter set with a comparatively soft EoS is observed to be a better choice to describe the sub-barrier nuclear fusion dynamics of reactions involving the Sn-isotopes.
\end{abstract}
\maketitle

\section{INTRODUCTION}
The study of the equation of state (EoS) of nuclear matter is imperative to shed light on numerous astrophysical events ranging from characteristics of neutron stars to core-collapse supernovae \cite{roca18,yasin20,ozel16,bogd19,bogd19a}. Understanding EoS is also crucial to probing the phase transitions in heavy-ion collisions. As a result, a considerable amount of theoretical as well as experimental efforts are being devoted to exploring the nuclear matter EoS \cite{roca18,yasin20,ozel16,bogd19,bogd19a}. The modulus of compression of nuclear matter, which is also known as nuclear incompressibility, is a fundamental quantity in EoS and plays an important role in the description of the physics of neutron stars, as well as in the bulk properties of finite nuclei \cite{peik07,li07,cao12,khan07,khan09,howa20}. The incompressibility of nuclear matter is interpreted as the curvature of the EoS of nuclear matter at the saturation point and is significant in understanding the dynamics of the nuclear system under small fluctuations in density \cite{peik07,li07,howa20}. There are no direct experimental techniques to deduce the value of the incompressibility of nuclear matter. However, the isoscalar giant monopole resonance (ISGMR), which is often termed the \textquote{breathing mode} of finite nuclei, is related to the behavior of a nucleus under small density fluctuations and provides an indirect experimental method to constrain the value of nuclear matter incompressibility \cite{cao12,hara01,garg18,blai95}. Thus, reliable experimental measurements of ISGMR energies for finite nuclei are preliminary in evaluating the incompressibility of infinite nuclear matter. A detailed description of the method used to determine the value of nuclear matter incompressibility using the ISGMR data can be found in Refs. \cite{cao12,hara01,garg18,blai95}.

The experimental ISGMR data of the heaviest doubly magic $^{208}$Pb nucleus has served as an optimal tool to investigate the nuclear matter incompressibility within various relativistic and non-relativistic approaches \cite{peik07,cao12,khan09,howa20,hara01, garg18, blai95,li08}. Furthermore, other magic nuclei such as $^{90}$Zr and $^{144}$Sm have also been used along with $^{208}$Pb. Various theoretical studies involving relativistic as well as non-relativistic interactions have led to the consensus that the range of nuclear matter incompressibility is 230 $\pm$ 40 MeV \cite{garg18,shlomo06,li21,colo14,khan12,marg18,dutra14}. In general, the procedure to extract nuclear matter incompressibility value from ISGMR data is understood to be independent of the choice of finite nuclei \cite{cao12,hara01,garg18,blai95}. However, recent experimental measurements for ISGMR data for the Tin isotopic chain ($^{112-124}$Sn) are found to be inconsistent with the established theoretical value of nuclear matter incompressibility from the ISGMR data of the $^{208}$Pb nucleus \cite{peik07,li07,howa20}. Models that accurately predict the ISGMR energies of the $^{208}$Pb nucleus are found to overestimate those of the $^{112-124}$Sn isotopes. Thus, the Sn-isotopes appear to be \textquote{compressible} or \textquote{soft} compared to the Pb, Sm and Zr nuclei \cite{peik07,li07,howa20}. The isotopes of cadmium (Cd) and molybdenum (Mo) were also observed to retain a similar compressible nature \cite{howa20,patel12}. This anomalous soft behavior of Sn isotopes has been a central topic of many theoretical and experimental studies. Consequently, there are several possible explanations for this incomprehensible soft behavior of Sn isotopes, such as the effects of mutually enhanced magicity (MEM) \cite{khan07} and superfluid pairing \cite{khan09,li08} on nuclear incompressibility. However, the MEM effects were disproved by experimental probes of the ISGMR energies of $^{204,206,208}$Pb isotopes \cite{patel13}, and the effects of pairing on ISGMR were found to be insufficient to address this softness \cite{khan09}. Thus, the question remains an open problem in nuclear structure physics, since neither the relativistic nor non-relativistic models can simultaneously fit the experimental ISGMR data of the Sn and Pb isotopes \cite{howa20}.

The low-energy heavy-ion reactions serve as an efficient tool for elucidating the correlation between the nuclear structure and reaction dynamics. The probability of nuclear fusion at energies around and below the Coulomb barrier, which is formed due to the strong interplay between the attractive nuclear and repulsive Coulomb interactions, is sensitive to various factors such as the nuclear shapes and orientations, nuclear shell effects, nuclear matter incompressibility, mass, charge and isospin asymmetry \cite{jiang21,back14,gautam17,pengo83,tora17,gupta06,mont17,toub17,das98,raj20,canto20,ghar22}. Moreover, coupling to the low-energy surface vibration states and neutron transfer also plays a crucial role in the description of the fusion dynamics at the sub-barrier energies \cite{jiang21,hagino98,deni20,zagr03,kohley13,zhen19,jones22,jiang15,liang07}. Consequently, considerable effort has been devoted to understanding the role of these nuclear structure properties on the reaction mechanism. For instance, in refs. \cite{ghar22,misicu07,misi06,esbe14}, the role of nuclear incompressibility is explored on the heavy-ion fusion cross-section at sub-barrier energies through the inclusion of a repulsive core in the nuclear potential. The formulation of the nuclear interaction potential formed between the fusing nuclei is essential to explore the impact of various factors of the entrance channel on the fusion process.  In our previous studies, the nuclear potential obtained within the double folding approach furnished with the self-consistent relativistic mean-field (RMF) formalism has become quite successful in describing the fusion dynamics of various heavy-ion reactions \cite{bhuy18,rana22,bhuy22}. Recently, the impact of nuclear shape degrees of freedom and orientations of the target nucleus was also included in the description of the nuclear potential within the RMF formalism \cite{rana23}. Moreover, the RMF formalism is also well-adopted to study the various structural properties of finite nuclei, as well as the characteristics of infinite nuclear matter, including nuclear incompressibility \cite{dutra14,meng16,lala09,ring96,biswal15,biswal14,vret05}. Following this, we aim to explore the effects of the above-discussed softness of Sn-isotopes on the nuclear fusion dynamics within the RMF formalism. For this, we have considered the even-even isotopes from  $^{112-124}$Sn chain exhibiting the anomalous soft behaviour along with the doubly magic $^{132}$Sn and $^{208}$Pb nuclei as targets with $^{64}$Ni as the projectile. The fusion cross-section for all the considered reactions is obtained using the extended $\ell-$summed Wong model \cite{kuma09,wong73}. The theoretical results are compared with the available experimental data \cite{jiang15,liang07,bock82,lesko86} to investigate the correlation of the peculiar soft behaviour of the Sn isotopes discussed above with the heavy-ion fusion cross-section. \\ 
The rest of the paper is structured as follows: Sec. \ref{theory} consists of a brief description of theoretical formalism adopted in the present analysis. The detailed discussion of the results obtained is provided in Sec. \ref{rslts} and Sec. \ref{smry} contains the summary and conclusions of the present study.

\section{THEORETICAL FORMALISM}
\label{theory}
The probability of nuclear fusion depends upon various structural properties of the interacting nuclei. The interaction potential formed between the fusing target and projectile nuclei is of fundamental essence to exploring the nuclear structure effects on the fusion dynamics. The total interaction potential between a spherical projectile and deformed target can be written as, 
\begin{eqnarray}
V_T(R,\beta_2,\theta_2)&=&V_C(R,\beta_2,\theta_2)+V_n(R,\beta_2,\theta_2)+\frac{\hbar^2\ell(\ell+1)}{2\mu R^2}. \nonumber \\
\label{vtot}
\end{eqnarray}
Here, $\theta_2$ denotes the orientation angle between the symmetry axis of the quadrupole deformed target and the inter-nuclear separation vector ($R$). $V_C(R,\beta_2,\theta_2)$ is the deformation and orientation dependent Coulomb potential \cite{wong73} and $\mu$ symbolizes the reduced mass of the target-projectile system. The values of quadrupole deformations ($\beta_2$) for all the target nuclei are taken from the experimental data given in \cite{raman03}. The term $V_n(R,\beta_2,\theta_2)$ in Eq. (\ref{vtot}) denotes the short range and attractive nuclear potential and is calculated within the well-known double folding approach \cite{satc79} as,
\begin{eqnarray}
V_{n}(\vec{R},\beta_2,\theta_2) &=& \int\rho_{p}(\vec{r}_p)\rho_{t}(\vec{r}_t(\beta_2,\theta_2))\nonumber \\
&& V_{eff}
\left( |\vec{r}_p-\vec{r}_t +\vec{R}| {\equiv}r \right) d^{3}r_pd^{3}r_t.
\label{fold}
\end{eqnarray}
Here, $\rho_p(\vec{r}_p)$ and $\rho_t(\vec{r}_t(\beta_2,\theta_2))$ are the total densities (sum of the proton and neutron densities) of the spherical projectile and quadrupole deformed target nuclei, respectively.  $V_{eff}$ symbolizes the effective nucleon-nucleon (NN) interaction potential. The self-consistent relativistic mean-field (RMF) formalism \cite{dutra14,bhuy18,meng16,lala09,ring96,biswal15,biswal14,vret05,sing12,sahu14} is adopted here to obtain the nuclear density distributions and microscopic effective NN interaction potential. The phenomenological RMF Lagrangian density describing the interaction between point-like nucleons through the exchange of mesons and photons \cite{dutra14,bhuy18,meng16,lala09,ring96,biswal15,biswal14,vret05,sing12,sahu14} can be written as,
 \begin{eqnarray}
{\cal L}&=&\overline{\psi}\{i\gamma^{\mu}\partial_{\mu}-M\}\psi +{\frac12}\partial^{\mu}\sigma
\partial_{\mu}\sigma \nonumber \\
&& -{\frac12}m_{\sigma}^{2}\sigma^{2}-{\frac13}g_{2}\sigma^{3} -{\frac14}g_{3}\sigma^{4}
-g_{\sigma}\overline{\psi}\psi\sigma \nonumber \\
&& -{\frac14}\Omega^{\mu\nu}\Omega_{\mu\nu}+{\frac12}m_{\omega}^{2}\omega^{\mu}\omega_{\mu}
-g_{w}\overline\psi\gamma^{\mu}\psi\omega_{\mu} \nonumber \\
&&-{\frac14}\vec{B}^{\mu\nu}.\vec{B}_{\mu\nu}+\frac{1}{2}m_{\rho}^2
\vec{\rho}^{\mu}.\vec{\rho}_{\mu} -g_{\rho}\overline{\psi}\gamma^{\mu}
\vec{\tau}\psi\cdot\vec{\rho}^{\mu}\nonumber \\
&&-{\frac14}F^{\mu\nu}F_{\mu\nu}-e\overline{\psi} \gamma^{\mu}
\frac{\left(1-\tau_{3}\right)}{2}\psi A_{\mu}.
\label{lag}
\end{eqnarray}
Here, $\psi$ is the Dirac spinor for nucleons of mass M, which interact through the exchange of $\sigma$, $\omega$, and $\rho$ mesons of masses  $m_\sigma$, $m_\omega$ and $m_\rho$, respectively. The terms $g_\sigma$, $g_\omega$, and $g_\rho$ signify the nucleon-meson coupling constants for the respective mesons and $g_2$, $g_3$ take into account the self-interaction properties of scalar $\sigma-$ mesons. The terms $\tau$ and $\tau_3$ in Eq. (\ref{lag}) denote the isospin and its third component, respectively, while $\Omega^{\mu\nu}$, $\vec B^{\mu\nu}$ and $F^{\mu\nu}$ are the field tensors for $\omega$, $\rho$ and photons, respectively.  The mass of $\sigma-$ mesons and the linear and non-linear coupling constants of mesons are known as the parameters of RMF formalism and are fine-tuned to fit the bulk properties of some magic shell nuclei as well as the properties of infinite nuclear matter. In the present analysis, we have adopted the non-linear NL3$^*$ parameter set, which successfully reproduces the experimental ISGMR energies of doubly magic $^{208}$Pb nucleus and gives the nuclear matter incompressibility value (K=258.25 MeV), which lies within its present acceptable range \cite{lala09}. Moreover, the nuclear densities and R3Y NN potential obtained for NL3$^*$ parameter are also observed to provide a satisfactory description of fusion dynamics of various heavy-ion reactions \cite{bhuy18,bhuy22} and references therein. We have also considered the hybrid parameter set \cite{piek09}, which is observed to provide a satisfactory description of the isoscalar monopole strengths of Sn-isotopes and yield K=230.01 MeV. It is worth noting here that this hybrid parameter set was constructed as a \textquote{test} model \cite{piek09} having the same $K$ value as the FSUGold parameter set \cite{rutel05} while yielding the other properties of nuclear matter (i.e. symmetry energy, saturation density, and energy per particle) similar to the NL3 parameter set \cite{lala97}. In addition, calculations are also performed with the set of NL1 parameters \cite{rein86}, which gives a relatively soft equation of state (EoS) with K=211.09 MeV. More details of the RMF parameterizations and field equations can be found in Refs. \cite{dutra14,bhuy18,meng16,lala09,ring96,biswal15,biswal14,vret05,sing12,sahu14} and references therein.
 
The effective nucleon-nucleon interaction potential ($V_{eff}$) has been obtained by solving the RMF equations for mesons within the limit of one-meson exchange \cite{bhuy18,sing12,sahu14}. This relativistic effective NN potential is known as the R3Y NN potential \cite{bhuy18,sing12,sahu14} and is written as,  \\
 \begin{eqnarray}
V_{eff}^{R3Y}(r)=\frac{g_{\omega}^{2}}{4{\pi}}\frac{e^{-m_{\omega}r}}{r}
+\frac{g_{\rho}^{2}}{4{\pi}}\frac{e^{-m_{\rho}r}}{r}
-\frac{g_{\sigma}^{2}}{4{\pi}}\frac{e^{-m_{\sigma}r}}{r} \nonumber \\
+\frac{g_{2}^{2}}{4{\pi}} r e^{-2m_{\sigma}r}
+\frac{g_{3}^{2}}{4{\pi}}\frac{e^{-3m_{\sigma}r}}{r}
+J_{00}(E)\delta(r). 
\label{r3y}
\end{eqnarray}
Here, the term $J_{00}(E)\delta(r)$ is a pseudo-potential that accounts for the long-range one-pion exchange potential (OPEP). Pairing correlations are considered within the BCS approach, and a blocking procedure is used to treat odd-mass nuclei \cite{zeng83,moli97,zhang11,hao12,lala99,lala99a,doba84,madl88,patra11}. The spherically symmetric nuclear densities for the projectile and target nuclei are also obtained from the RMF Lagrangian density (see Eq. \ref{lag}). The impact of nuclear shape degrees of freedom and orientation is further included through the radius vector in the spherical symmetric target densities obtained within RMF formalism for different parameter sets discussed above. The nuclear radius of an axially deformed nucleus can be written in terms of spherical harmonic expansion as \cite{moll16, bohr52,bohr53}, 
\begin{eqnarray}
{r}_t(\beta_2,\theta_2)={r}_{0t}[1+\sqrt{(5/4\pi)}\beta_2 P_2(cos\theta_2)].
\label{drad}
\end{eqnarray}
Here, $r_{0t}$ is a symmetric spherical radial vector. The deformed densities obtained using Eq. \ref{drad} along with the relativistic R3Y NN potentials are further used to calculate the nuclear interaction potential using Eq. \ref{fold}. Characteristics of the total interaction potential (see Eq. \ref{vtot}), i.e. barrier height ($V_B^\ell$), barrier position ($R_B^\ell$) and barrier curvature ($\hbar\omega_\ell$) are further used to evaluate the nuclear fusion probability. In the literature, many analytical expressions have been developed to determine the probability of penetration of the barrier to avoid tedious numerical evaluation \cite{jiang21,toub17,canto20,hill53}. The Hill-Wheeler formula obtained using the parabolic barrier approximation \cite{hill53} is one of the widely adopted approaches that has become quite successful in determining the probability of heavy-ion fusion at around and above the barrier energies, especially for reactions involving intermediate and heavy-mass nuclei \cite{jiang21,canto20,bhuy18,rana22,bhuy22}. As the present analysis also focuses on reactions involving nuclei from the same mass regions, we have also used the Hill-Wheeler approximation of the parabolic barrier to calculate the penetrability ($P_{\ell}$) as 
 \begin{eqnarray}
P_\ell(E_{c.m},\theta_2)=\Bigg[1+exp\bigg(\frac{2 \pi (V_{B}^{\ell}(\theta_2)-E_{c.m.})}{\hbar \omega_{\ell}(\theta_2)}\bigg)\Bigg]^{-1}. 
\end{eqnarray}
Here, $E_{c.m.}$ is the energy of the target-projectile system in the center of the mass frame. Finally, the fusion and/or capture cross-section is obtained using the extended version of the Wong formula entitled the $\ell-$summed Wong model \cite{bhuy18,kuma09,wong73}. In the $\ell-$summed Wong model, the actual angular momentum dependence of interaction potential is taken into account, and the cross-section is written in terms of $\ell-$partial waves \cite{bhuy18,kuma09,wong73} as, 
\begin{eqnarray}
\sigma(E_{c.m.},\theta_2)=\frac{\pi}{k^{2}} \sum_{\ell=0}^{\ell_{max}}(2\ell+1)P_\ell(E_{c.m},\theta_2).
\label{crs}
\end{eqnarray}
Here, $k=\sqrt{\frac{2 \mu E_{c.m.}}{\hbar^{2}}}$. The $\ell-$values are obtained using the sharp cut-off model \cite{beck81} at the above barrier energies. These $\ell_{max}$-values are the same as the critical angular momenta (which is also symbolized as $\ell_c$ or $\ell_{cr}$) for complete fusion \cite{beck81}. The sharp cut-off model is only applicable at above barrier energies, so an energy-dependent extrapolation is used to obtain the $\ell_{max}$-values at below barrier energies. As present work mainly aims to explore the impact of nuclear incompressibility and the softness of Sn isotopes on the characteristics of the fusion barrier obtained from different sets of RMF parameter, therefore the effects of channel coupling are not taken into account in the cross section calculations. The $\ell-$summed cross-section is obtained using Eq. (\ref{crs}) at different target orientation angles ($\theta_2 = 0$ to $\pi/2$). Moreover, the symmetry axis of the deformed target nucleus is not frozen at a particular angle during the nuclear collision, so the integrated cross-section over the target orientation angle ($\theta_2$) is obtained. This method to calculate the total cross-section is well-adopted in various studies of nuclear fusion \cite{kuma09,long08,lari16,rash96, arit12,sun23}. For the case of spherical projectile and deformed target, the total integrated cross-section can be written as, 
\begin{eqnarray}
\sigma_{int}(E_{c.m.})=\int_{0}^{\pi/2} \sigma(E_{c.m.},\theta_2) sin\theta_2 d\theta_2.
\label{icrs}
\end{eqnarray}
This theoretical approach to calculate the cross-section for reactions involving axially deformed targets within the $\ell-$summed Wong model supplemented with RMF formalism is used to describe of fusion dynamics of reactions involving even-even Sn isotopes, which are observed to exhibit an incomprehensible soft nature. 

\section{RESULTS AND DISCUSSION}
\label{rslts}
This section aims to explore the effects of nuclear matter incompressibility and the anomalous compressible behavior of Sn isotopes on the dynamics of nuclear fusion. For this, we have considered the reaction systems involving the so-called soft even-even isotopes in $^{112-124}$Sn chain as targets and $^{64}$Ni as the projectile. The results of these fusion reactions are compared with those that contain doubly magic $^{132}$ Sn and $^{208}$ Pb nuclei as targets along with the same projectile, i.e., $^{64}$Ni. As mentioned above, the impact of target quadruple deformations is taken into account in the calculations of nuclear interaction potential formed between the fusing nuclei. The main ingredient of this total interaction potential is the attractive and short-range nuclear potential, which is calculated here in terms of nuclear densities and effective NN interaction using the double-folding approach. The well-established relativistic mean-field (RMF) formalism is adopted to determine the microscopic R3Y effective NN potential and nuclear density distributions. Here, the non-linear RMF parameter sets NL3$^*$ \cite{lala09}, hybrid \cite{piek09} and NL1 \cite{rein86} are used, which are observed to give different values of incompressibility of isospin symmetric nuclear matter. In our previous studies \cite{bhuy18,bhuy22}, the NL3$^*$ (K=258.25) parameter set is also observed to give a reasonable fit to the fusion and/or capture cross-section of various heavy-ion reactions. The hybrid parameter, which produces lower values of nuclear matter incompressibility (K=230.01 MeV) and symmetry energy \cite{piek09} than the NL3$^*$ \cite{lala09} parameter set, is observed to give a satisfactory fit to the experimental giant monopole resonance (GMR) energies of the Sn isotopes. In addition, the hybrid parameter set and NL3$^*$ models are observed to give almost similar values of other characteristics of nuclear matter such as saturation density, and energy per particle. In addition to these, we have also considered the NL1 parameter set \cite{rein86}, which gives a very soft equation of state (EoS) with K = 211.09 MeV compared to NL3$^*$ and hybrid models. 
\begin{figure*}
   \centering
    \includegraphics[scale=0.22]{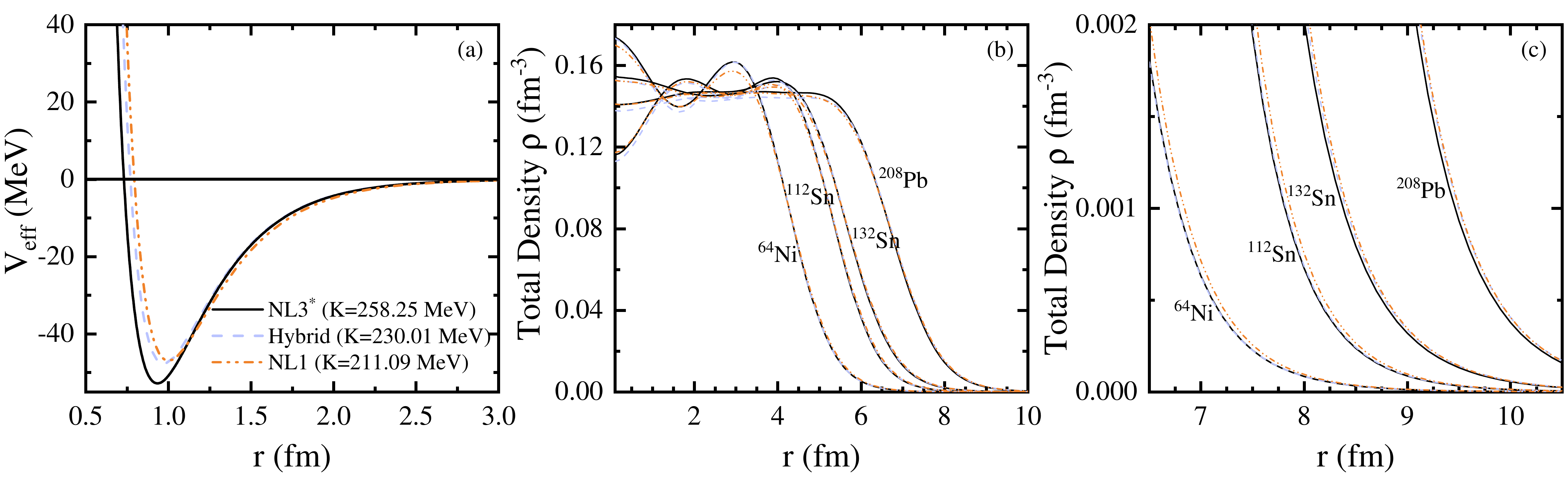}
    \caption{(a) The effective R3Y NN potential, (b) radial distribution of total nuclear densities (c) total nuclear densities at surface region calculated using NL3$^{*}$ (solid black line), hybrid (dashed blue lines) and NL1 (dash double dotted orange lines) parameter sets and See text for details.}
    \label{fig1}
\end{figure*}
Figure \ref{fig1} shows the effective NN potential and spherically symmetric nuclear density distributions obtained within the RMF approach for different sets of parameters. Fig. \ref{fig1}(a) shows the R3Y NN potential calculated for NL3$^*$ (solid black line), hybrid (dashed blue line) and NL1 (dashed double dotted orange line) parameter sets. The different RMF parameter sets are observed to show different depths for the attractive core of the effective nucleon-nucleon (NN) potential. The NL3$^*$ parameter set with a comparatively stiffer EoS is observed to give the most attractive effective NN interaction at the lower inter-nucleon separation (r), whereas the NL1 parameter set with soft EoS gives the most attractive NN potential at the higher inter-nucleon separation ($r \ge 1.2$ fm). Fig. \ref{fig1}(b) shows the radial distribution of total density (sum of proton and neutron densities, $\rho=\rho_P+\rho_N$) obtained within the different RMF parameter sets for the representative cases of light mass projectile $^{64}$Ni, $^{112}$Sn and $^{132}$Sn isotopes with lower and higher $N/Z$ ratio, respectively and the heavy doubly magic $^{208}$Pb nucleus. The density of light mass $^{64}$Ni shows a decrease in density in the core region due to the combined effects of the shell effects and Coulomb repulsion \cite{rein02,afan05,chu10}. The density of the heavy $^{208}$Pb nucleus is observed to show a comparatively flat curve extended toward the higher radial region. By comparing the densities of $^{112,132}$Sn isotopes at the surface region, it is observed that the density increases with the increase in neutron number. As nuclear fusion is a surface phenomenon, the tail region of the nuclear densities plays the most crucial role in the description of nuclear fusion \cite{gupta07,rana22}. Fig. \ref{fig1}(c) shows the magnified view of the total density distribution in the surface region obtained using different parameter sets under study. Comparing the densities obtained for the NL3$^*$ and hybrid parameter sets, it is observed that the densities in the surface region increase with a decrease in the nuclear incompressibility value. Moreover, the NL1 parameter set with softer EoS is observed to give the highest total density in the surface region for all nuclei studied. The small difference in the density at the surface region further influences the fusion probability. Further, the effects of nuclear deformations are introduced through the radius vector (see Eq. \ref{drad}) in the RMF densities for the target nuclei. These deformed densities and relativistic R3Y effective NN potential are further employed to obtain the nuclear interaction potential for all the considered reactions.
\begin{figure*}
    \centering
    \includegraphics[scale=0.22]{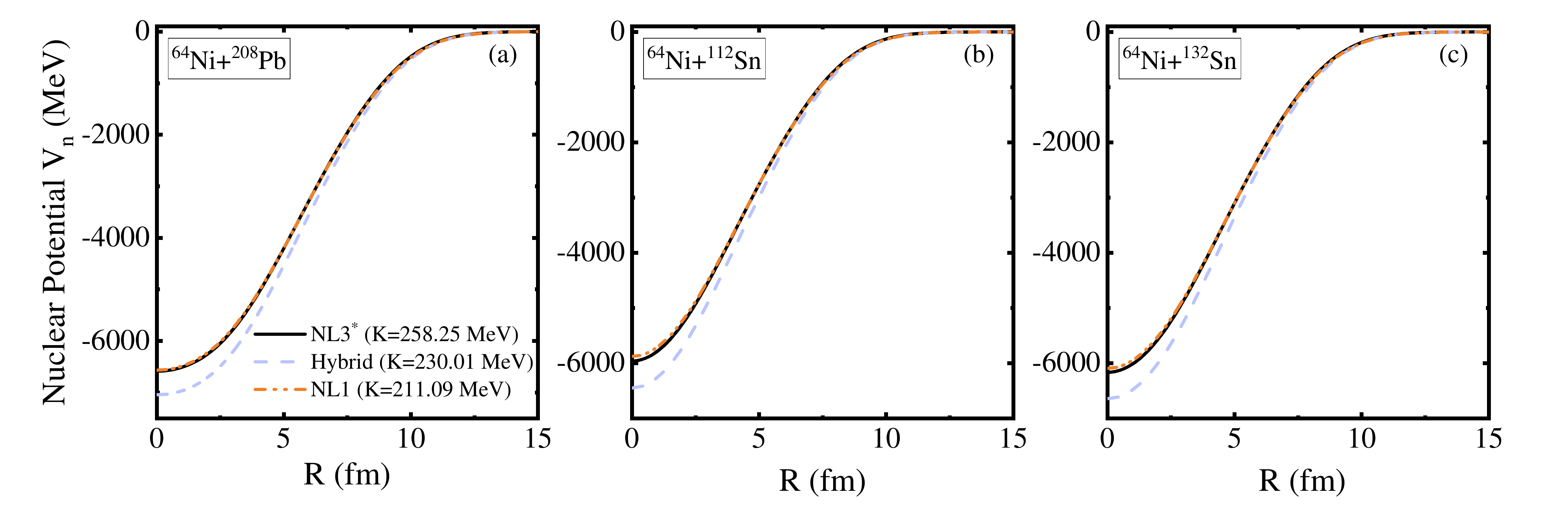}
    \caption{Nuclear interaction potential $V_n$ (MeV) for obtained using different RMF parameter sets as a function of inter-nuclear separation R (fm) for (a) $^{64}$Ni + $^{208}$Pb and (b) $^{64}$Ni+$^{112}$Sn and $^{64}$Ni+$^{132}$Sn reactions.  }
    \label{fig2}
\end{figure*}


\begin{table*}[]
\centering
\caption{The positions $R_B$ (in fm) and heights $V_B$ (in MeV) of the fusion barriers obtained using NL3$^{*}$ (K=258.25 MeV), hybrid (K=230.01 MeV) and NL1 (K=211.09 MeV) parameter sets for all the reactions under study.}
\begin{tabular}{ccccccc}
\hline
\multirow{2}{*}{Reaction} & \multicolumn{2}{c}{NL3$^*$} & \multicolumn{2}{c}{Hybrid} & \multicolumn{2}{c}{NL1} \\
                          & $R_B$ (fm)         & $V_B$ (MeV)           &    $R_B$ (fm)           &     $V_B$ (MeV) &         $R_B$ (fm)          & $V_B$ (MeV)          \\
                          \hline
$^{64}$Ni+$^{208}$Pb                & 13.4       & 233.51      & 13.5        & 232.31       & 13.5      & 231.54     \\                          
$^{64}$Ni+$^{112}$Sn                 & 12.1       & 157.68      & 12.1        & 157.16       & 12.2      & 156.09      \\
$^{64}$Ni+$^{114}$Sn                 & 12.1       & 156.89      & 12.2        & 156.34       & 12.2      & 155.32      \\
$^{64}$Ni+$^{116}$Sn                & 12.2       & 155.96      & 12.2        & 155.37       & 12.3      & 154.44      \\
$^{64}$Ni+$^{118}$Sn               & 12.3       & 155.06      & 12.3        & 154.49       & 12.4      & 153.59      \\
$^{64}$Ni+$^{120}$Sn                & 12.3       & 154.04      & 12.4        & 153.42       & 12.4      & 152.61      \\
$^{64}$Ni+$^{122}$Sn                & 12.4       & 153.60      & 12.4        & 152.97       & 12.5      & 152.18      \\
$^{64}$Ni+$^{124}$Sn                & 12.4       & 153.17      & 12.5        & 152.52       & 12.5      & 151.75      \\
$^{64}$Ni+$^{132}$Sn                & 12.6       & 151.47      & 12.6        & 150.79       & 12.7      & 150.05      \\
\hline
\end{tabular}
\label{tab1}
\end{table*}

First, we analyze the nuclear potential calculated by taking both projectile and target nuclei to be spherically symmetric. Fig. \ref{fig2} shows the nuclear potential $V_n$ (MeV) for spherical case (i.e. considering $\beta_2=0.0$) as a function of the nuclear separation distance $R$ (fm) for the illustrative cases of (a) $^{64}$Ni + $^{208}$Pb and (b) $^{64}$Ni+$^{112}$Sn and (c) $^{64}$Ni+$^{132}$Sn reactions. It is noticed from Fig. \ref{fig2} that the hybrid parameter set gives the most attractive nuclear potential at smaller inter-nuclear separation. However, at larger inter-nuclear separation, which plays the most crucial role in the nuclear fusion mechanism, the NL1 parameter set with a soft EoS (K=211.09 MeV) gives the most attractive nuclear potential. Moreover, the nuclear potential at larger $R$ becomes repulsive with the increase in the nuclear matter compressibility (K) value of the parameter set used to obtain the nuclear densities and R3Y effective NN potential. The resultant of the short-range nuclear and long-range Coulomb potentials give rise to the fusion barrier. The characteristics of this fusion barrier such as its height ($V_B$), position ($R_B$) and frequency play the most important role in determining the heavy-ion fusion cross-section, especially at the sub-barrier energy region. The positions $R_B$ (in fm) and heights $V_B$ (in MeV) of the fusion barriers obtained using NL3$^{*}$ (K=258.25 MeV), hybrid (K=230.01 MeV) and NL1 (K=211.09 MeV) parameter sets for all the considered reactions are given in Table \ref{tab1}. It can be observed from Table \ref{tab1} that the height of the fusion barrier decreases as we move from the NL3$^*$ parameter set with a higher value of nuclear matter incompressibility (K = 258.25 MeV) to the set of hybrid parameters having lower nuclear matter incompressibility (K=230.01 MeV) at saturation. The barrier height is observed to decrease further for the NL1 parameter set that has a soft EoS with K=211.09 MeV. Further, it can be noticed from Table \ref{tab1}  that the height of the fusion barrier decreases with increasing neutron number (N) of Sn-isotopes. This is because, with an increase in the number of neutrons of Sn isotopes, the nuclear potential becomes more and more attractive, whereas the charge-dependent repulsive Coulomb potential remains the same. This infers that a lower fusion barrier and, consequently, a higher cross-section will be obtained for the reactions involving neutron-rich nuclei.

\begin{figure}
\centering
\includegraphics[scale=0.4]{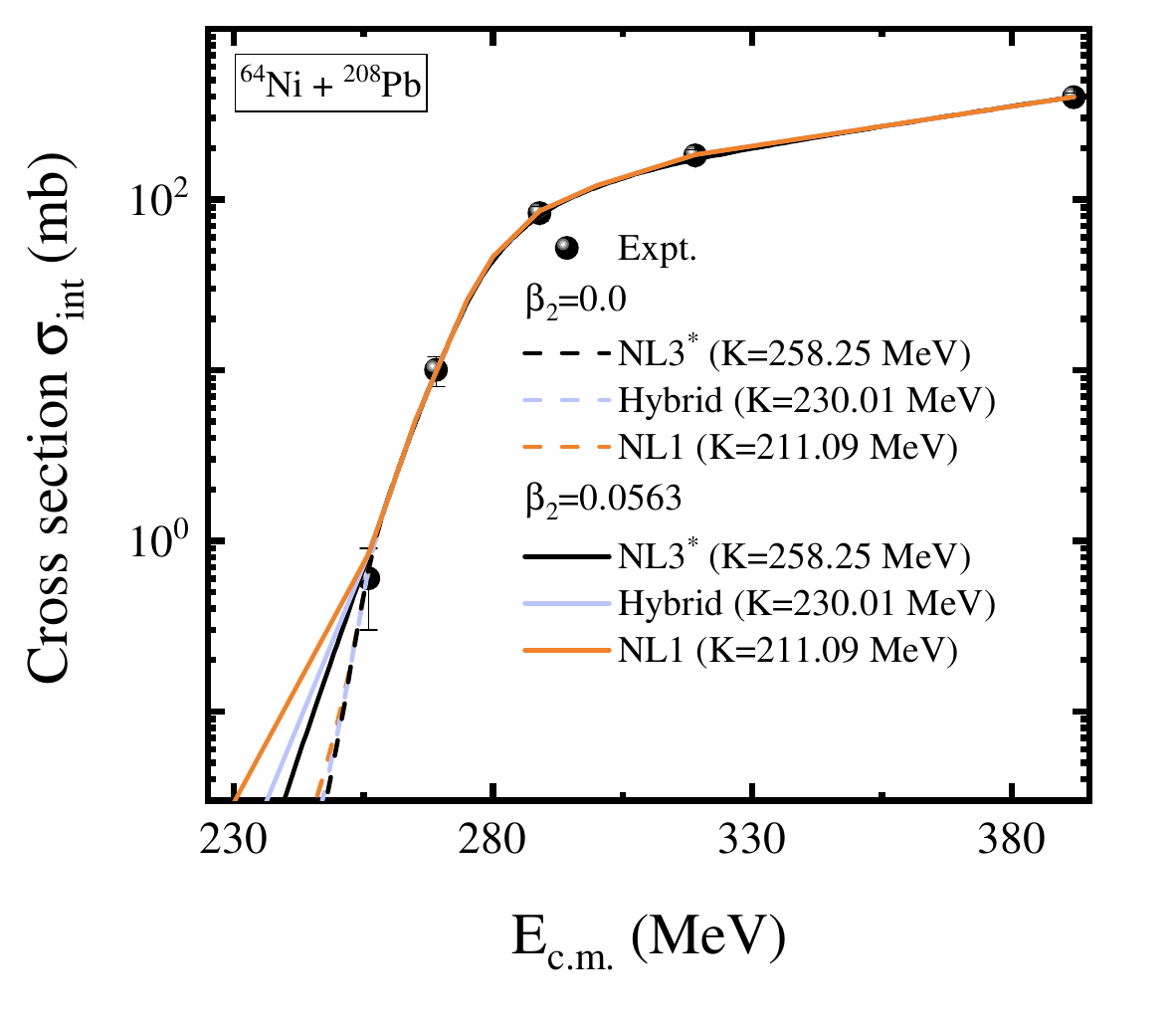}
\caption{The fusion cross section $\sigma_{int}$ (mb) calculated using NL3$^{*}$ (black), hybrid (blue) and NL1 (orange) parameter sets within the $\ell-$summed Wong model as a function of center of mass energy $E_{c.m.}$ (MeV) for $^{64}$Ni+$^{208}$Pb reaction. The dashed line signifies the cross-section calculated for the spherical case ($\beta_2=0$) and solid lines are for the cross-section obtained with the inclusion of target quadrupole deformations ($\beta_2>0$). The corresponding experimental data are taken from \cite{bock82}.}
   \label{fig3}
\end{figure}

It is worth noting here that barrier characteristics are shown for the case of spherical projectile and target nuclei in Table \ref{tab1}. It is well-known that the shape of interacting partners also affects the fusion cross-section at the sub-barrier energies. The deformation from the spherical symmetric shape of either or both reacting partners changes the interaction radius and hence the interaction potential. Following this, we have also included the effect of target quadrupole deformation ($\beta_2$) in calculations of nucleus-nucleus potential. On the inclusion of nuclear shape degrees of freedom, the orientation ($\theta_2$) of the axially deformed target with respect to the inter-nuclear separation vector also affects the interaction potential. Thus, the fusion barrier characteristics for all the considered reactions are calculated at each orientation angle ($\theta_2= 0$ to $\pi/2$). These deformation and orientation-dependent barrier characteristics are further utilized to obtain the barrier penetration probability and $\theta_2-$integrated cross-section. First, we analyze the capture cross-section for $^{64}$Ni+$^{208}$Pb reaction calculated using the $\ell-$summed Wong model supplemented with nuclear potential obtained using the three sets of parameters of the RMF having different values of incompressibility of nuclear matter between K = 211.09-258.25 MeV. The $\ell_{max}-$values are obtained from the sharp cut-off model \cite{beck81} at the above barrier centre of mass energies and are extrapolated for the below-barrier region. The cross-section $\sigma_{int}$ (mb) calculated without (dashed lines) and with (solid lines) the inclusion of target deformation for $^{64}$Ni+$^{208}$Pb reaction is shown in Fig. \ref{fig3} as a function of the centre of mass energy $E_{c.m.}$ (MeV). The experimental data taken from Ref. \cite{bock82} is also shown for comparison. It can be observed from Fig. \ref{fig3} that at below barrier energies, the NL1 parameter set with the lowest value of nuclear matter incompressibility, i.e., K=211.09 MeV, gives the highest cross-section. The below barrier cross-section decreases as the incompressibility of the RMF parameter set increases, with the NL3$^*$ set giving the lowest cross-section. These results infer that the capture cross-section at the below-barrier centre of mass energies increases with the decrease in nuclear matter incompressibility (K). Further, on analyzing the cross-section obtained with the inclusion of nuclear shape degrees of freedom, it is noticed that the sub-barrier cross-section increases on the inclusion of the impact of quadrupole deformation in the calculations of the interaction potential. Moreover, the theoretical cross-section obtained using different RMF parameter sets overlaps at the above barrier energies and also shows a nice agreement with the experimental data. This is because the effects of nuclear structure are suppressed at the above barrier energies and the angular momentum effects dominate.

\begin{figure*}
\centering
\includegraphics[scale=0.35]{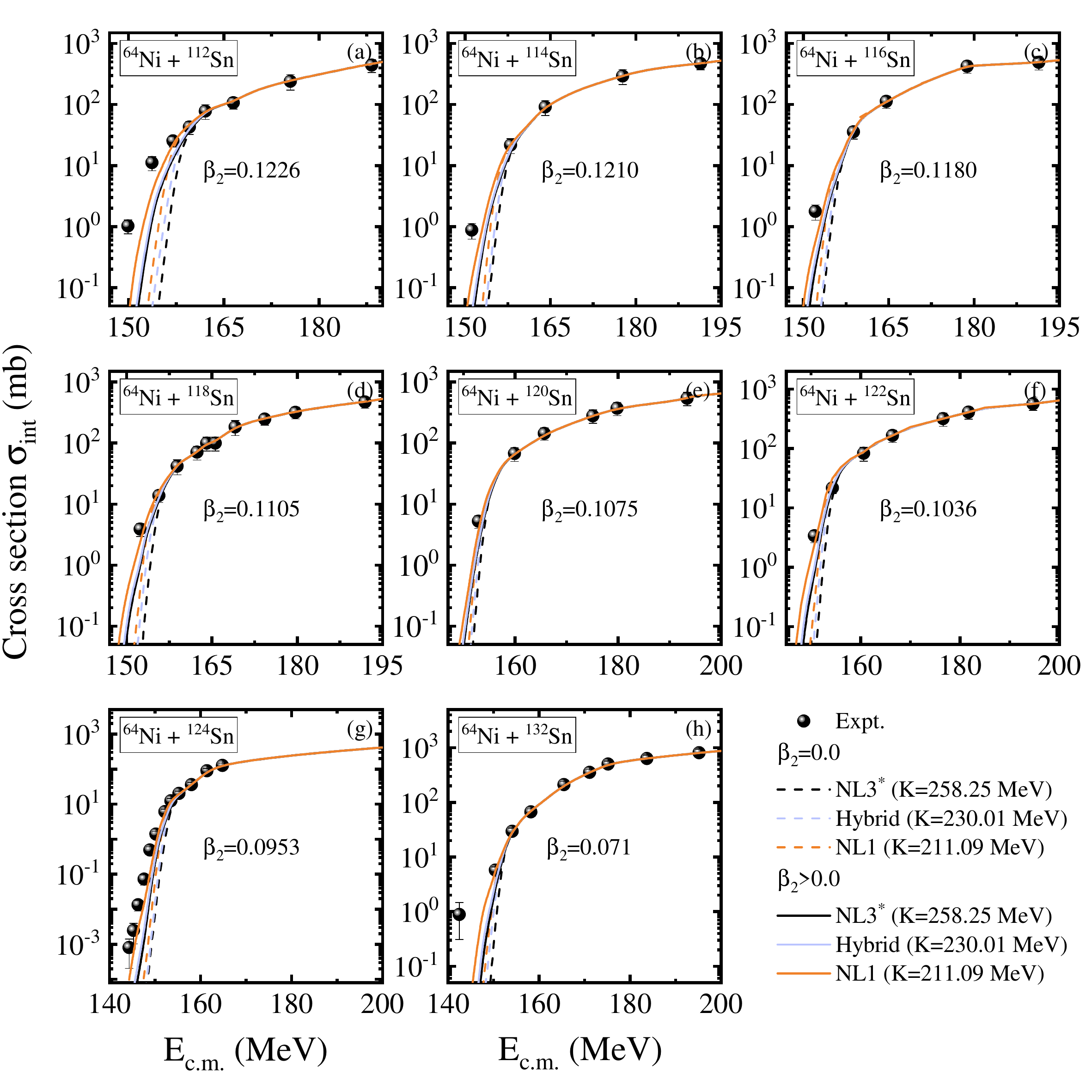}
\caption{The fusion cross section $\sigma_{int}$ (mb) calculated using NL3$^{*}$ (black), hybrid (blue) and NL1 (orange) parameter sets within the $\ell-$summed Wong model as function of center of mass energy $E_{c.m.}$ (MeV) for the even-even $^{64}$Ni+$^{112-124,132}$Sn systems. he dashed signifies the cross-section calculated for the spherical case ($\beta_2=0$) and solid lines are for the cross-section obtained with the inclusion of target quadrupole deformations ($\beta_2>0$). The corresponding experimental data is taken from \cite{jiang15,liang07,lesko86}.}
\label{fig4}
\end{figure*}

As mentioned above, the Sn-isotopes appear to be \textquote{soft} or \textquote{compressible} in comparison to the doubly magic $^{208}$Pb nucleus. To explore the effect of this peculiar nature of Sn-isotopes on the fusion mechanism, next, we have calculated the cross-section for the even-even $^{112-124,132}$Sn targets with the same $^{64}$Ni projectile. The calculated cross-section for 8 even-even reactions, namely $^{64}$Ni+$^{112-124,132}$Sn, is plotted in Fig. \ref{fig4} for all three considered RMF parameter sets along with the experimental data \cite{jiang15,liang07,lesko86}. The notations used in Fig. \ref{fig4} are similar to those adopted in Fig. \ref{fig3}. The $\beta_2$ values for considered even-even $^{64}$Ni+$^{112-124}$Sn nuclei are taken from the experimental data given in \cite{raman03}, whereas $\beta_2$ for $^{132}$Sn is taken from the NuDat3 database. It is noticed from Fig. \ref{fig4} that the cross-section calculated using the $\ell$-summed Wong model supplemented with the nuclear potentials obtained from RMF formalism shows a nice agreement with available experimental data at the above-barrier energies. However, at the below barrier energies, the cross-section obtained for NL3$^*$ (K=258.25 MeV) parameter set (black lines) underestimates the experimental cross-section for all the reactions involving even-even Sn-isotopes. The match between the experimental data and theoretical cross-section obtained for NL3$^*$ parameter set improves a little on the inclusion of nuclear shape degrees of freedom, but still, an underestimation remains at below barrier energies. It is worth noting here that in our previous studies \cite{bhuy18,rana22,bhuy22,rana23} and references therein, a comparatively better match with the experimental data was observed for NL3 (K=271.53 MeV) as its improved version i.e., NL3$^*$ (K=258.25 MeV) parameter set.  

The discrepancy between the calculated and experimental cross-section for reactions involving Sn-isotopes can be connected to the above-discussed softness of Sn-isotope. When correlating our results with the studies of isoscalar giant monopole resonance (ISGMR) studies for Sn-isotopes within the RMF formalism for different parameter sets \cite{biswal15,biswal14}, NL3$^*$ is observed to overestimate the experimental ISGMR energies for Sn-isotopes (see Table 2. in \cite{biswal14}). On the other hand, a better match with the experimental ISGMR data of $^{208}$Pb is observed with NL3$^*$ parameter set \cite{lala09,biswal14}. Moreover, the NL1 parameter set is observed to underestimate the ISGMR energies for Sn-isotopes as well for the $^{208}$Pb. In \cite{biswal15}, it is concluded that the parameters having nuclear incompressibility values between 210-230 MeV are suitable to reproduce the monopole energies of Sn-isotopes. Moreover, in \cite{piek09}, the non-linear hybrid RMF parameter set is proposed as a \textquote{test} model to address the overestimation of ISGMR data of even-even Sn-isotopes within the RMF formalism. This hybrid parameter gives the value of nuclear matter incompressibility (K = 230.01 MeV) similar to the FSUGold parameter set \cite{rutel05} and yields other nuclear matter properties such as symmetry energy, saturation density, and energy per particle similar to the well-known NL3 parameter set \cite{lala97}. Following these observations, we have also calculated the cross-section for reactions involving Sn-isotopes with the hybrid model (K=230.01 MeV), which is constructed to describe the ISGMR data of Sn-isotopes and also with NL1 parameter set (K = 211.09 MeV) having comparatively soft EoS. It can be noted from Fig. \ref{fig4} that the hybrid (blue lines) and NL1 (orange lines) with lower nuclear matter incompressibility values give higher cross-section at sub-barrier energies as compared to the NL3$^*$ parameter set. In other words, the cross-section is observed to increase moderately with the increase in the nuclear matter incompressibility value at the below barrier energies for all the reactions under study. Moreover, the nuclear potential obtained for the NL1 parameter set (solid orange lines) with the inclusion of target quadrupole deformations is observed to give a comparatively better fit to the cross-section of $^{64}$Ni+$^{116-122}$Sn but still underestimates the experimental cross-section for reactions involving other even-even Sn-isotopes. The discrepancy between the experimental and theoretical cross-section becomes more prominent as we move towards far below-barrier energy regions. 
\begin{figure}
\centering
\includegraphics[scale=0.35]{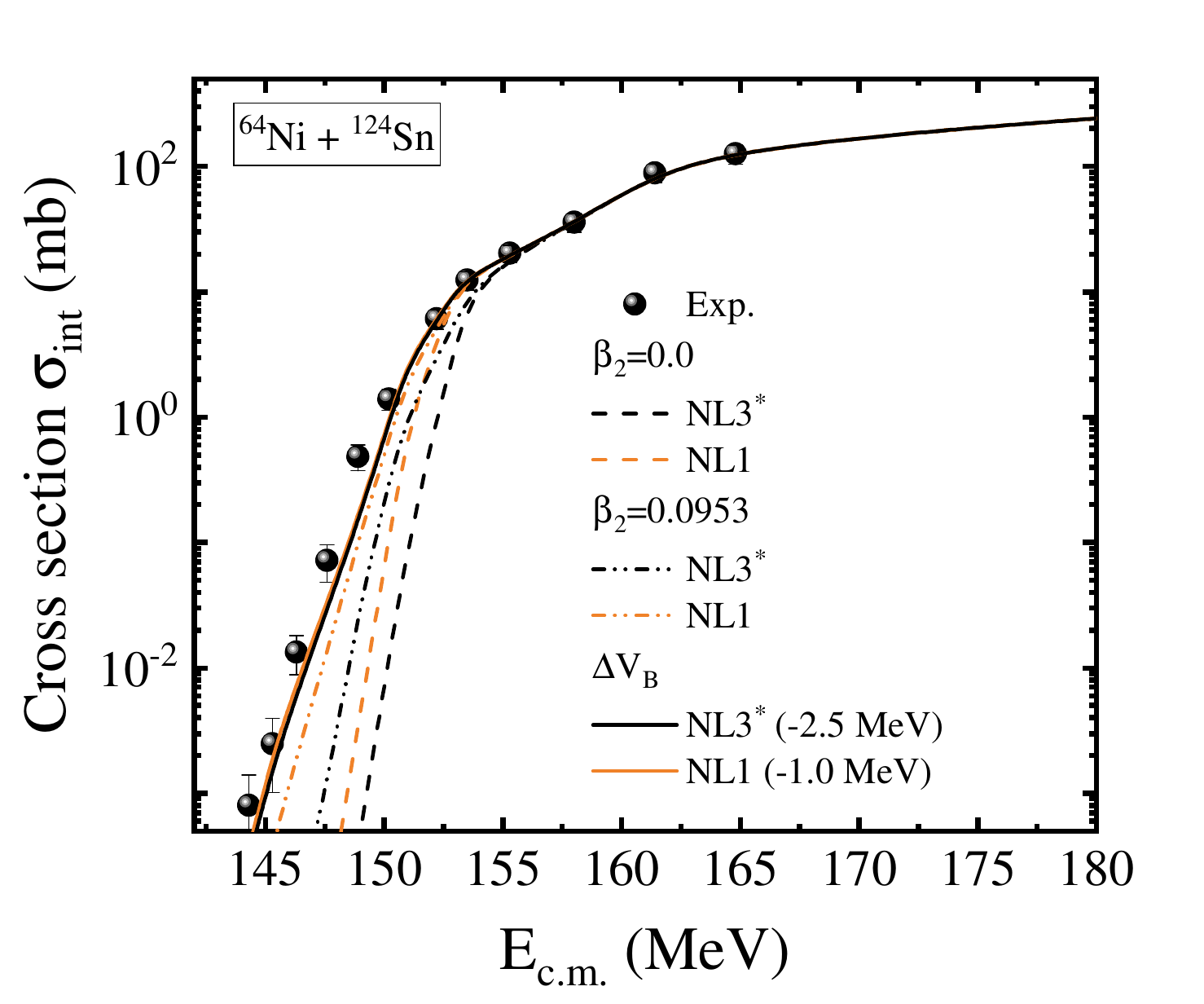}
\caption{The fusion cross section $\sigma_{int}$ (mb) calculated using NL3$^{*}$ (black) and NL1 (orange) parameter sets along with the barrier modification (solid lines) within the $\ell-$summed Wong model as a function of the centre of mass energy $E_{c.m.}$ (MeV) for the $^{64}$Ni+$^{124}$Sn reaction. The dashed lines signify the calculations done without the incorporation of nuclear deformations and dashed double-dotted lines signify the cross-section calculated with the inclusion of target deformation. The experimental data are taken from \cite{jiang15}.}
    \label{fig5}
\end{figure}

These observations of variation in fusion characteristics with different RMF parameter sets are in line with the theoretical investigations for $^{16}$O+$^{208}$Pb using different Skyrme forces associated with different values of nuclear incompressibility \cite{ghodsi16}. In \cite{ghodsi16} and in the present study, it is observed that the height of the fusion barrier increases with increasing incompressibility. Thus, the softness of the nucleus will decrease the height of the fusion barrier. This can also be understood in terms of nuclear radius. As per Fig. \ref{fig1}(a), more extended nuclear density distribution in the surface region is observed for the NL1 parameter set which gives a soft equation of state. From this, it can be inferred that the surface density of a soft or compressible nucleus will be more extended which will result in a larger nuclear radius. This small increment in surface density leads to the lower barrier height and higher cross-section for the NL1 parameter set in comparison to the other two considered parameter sets (hybrid and NL3$^*$) with a high value of nuclear matter incompressibility. Thus, the fusion cross-section increases with a decrease in the nuclear incompressibility value of the RMF parameter set. Furthermore, the nuclear interaction potential and hence the cross-section depends on various structural aspects of the fusing nuclei. In the present study, we have only included the impact of target quadrupole deformation in the calculations of microscopic nuclear potential obtained within the RMF formalism. However, other factors such as couplings to the low-energy surface vibration states and neutron transfer channels also affect the sub-barrier fusion cross-section. In order to account for the combined effects of the softness of Sn isotopes and the other effects of the nuclear structure, we have done the barrier modification for $^{64}$Ni+$^{124}$Sn reaction as experimental data at far sub-barrier energies is available for this reaction \cite{jiang15}. This approach of barrier modification, although implausible, has been used to account for various nuclear structural effects \cite{kuma09,vijay22}. Figure \ref{fig5} shows the cross-section $^{64}$Ni+$^{124}$Sn reaction with a barrier modification (solid lines) done in the barrier heights obtained using NL3$^{*}$ (black) and NL1 (orange) parameter sets along with nuclear shape degrees of freedom. It is noted that the NL1 parameter set with soft EoS having K= 211.09 MeV underestimates the cross-section at sub-barrier energies and a further barrier lowering by 1.0 MeV is needed to address the mismatch. On the other hand, for the NL3$^{*}$ parameters set which gives a higher value of nuclear matter incompressibility, a barrier modification of -2.5 MeV is required to give a better match with the experimental data. All these observations indicate that the softness of Sn-isotopes leads to the lowering of the fusion barrier and, consequently, to an enhancement in the fusion probability at below-barrier energies. In other words, the signature of the soft behaviour of Sn-isotopes is also observed in their fusion dynamics. The NL3$^*$ parameter set, which fails to explain the compressible nature of Sn-isotopes, is also observed to underestimate the cross-section for reactions involving these Sn-isotopes. Barrier modification is required to address the discrepancy between the experimental and calculated cross-section caused by the softness of Sn-isotopes. Moreover, the magnitude of the required barrier modification is significantly less for the RMF parameter set with soft EoS. On the other hand, no such barrier modification is needed to address the experimental data within the $\ell$-summed Wong model furnished with nuclear potential from the RMF formalism for the $^{64}$Ni+$^{208}$Pb reaction considered in the present analysis.

\section{SUMMARY AND CONCLUSIONS}
\label{smry}
The effects of nuclear matter incompressibility (K) and the soft nature of Sn-isotopes are probed on the heavy-ion fusion dynamics. For this, the cross-section for even-even $^{64}$Ni+$^{112-124,132}$Sn and $^{64}$Ni+$^{208}$Pb reactions is calculated within the $\ell-$summed Wong model equipped with nuclear potential from the relativistic mean-field (RMF) formalism with the inclusion of target quadrupole deformations for three non-linear parameter sets which yield different values for the isospin symmetric nuclear matter incompressibility (K) at saturation. First, the results of non-linear NL3$^*$ and hybrid parameter sets, which give almost similar values of nuclear matter properties except for the incompressibility at saturation and symmetry energy, are compared. It is observed that the cross-section at around the barrier energies increases on moving from the NL3$^*$ parameter set with K=258.25 MeV to the hybrid parameter set with K=230.01 MeV. Further, an increase in the cross-section is observed for the NL1 parameter set having a soft EoS with a lower value of nuclear incompressibility (K=211.09 MeV) \cite{dutra14}.

The calculated cross-section is also compared with the available experimental data, and a nice agreement is obtained for $^{64}$Ni+$^{208}$Pb reaction for all the considered non-linear RMF parameter sets at the above barrier energies. On the contrary, the cross-section obtained using the nuclear potential calculated within the RMF (NL3$^*$) approach is observed to underestimate the experimental data at below barrier energies for all the considered reactions involving even-even $^{112-124,132}$Sn isotopes. Further, an increase in the cross-section is observed for the hybrid (K=230.01 MeV) and NL1 (211.09 MeV) with lower values of nuclear matter incompressibility. A better match to the experimental cross-section for $^{64}$Ni+$^{112-124,132}$Sn reactions is observed for the NL1 parameter set. In correlating this mismatch between the theoretical and experimental data observed in the ISGMR studies of Sn-isotopes, it is noticed that the effect of the softness of Sn-isotopes also persists in their fusion dynamics. The soft or compressible nature of Sn-isotopes leads to the enhancement in the experimental cross-section at below-barrier energies and the RMF parameter set with comparatively soft EoS becomes a better choice to describe the fusion dynamics of reactions involving these Sn-isotopes.
 
\section*{Acknowledgements}
This work has been supported by the Science Engineering Research Board (SERB) File No. CRG/2021/001229, FAPESP Project Nos. 2017/05660-0, and FOSTECT Project No. FOSTECT.2019B.04.


\begin{thebibliography}{99}
\bibitem{roca18}
X. Roca-Maza, N. Paar, Prog. Part. Nucl. Phys. {\bf 101}, 96 (2018).
\bibitem{yasin20}
H. Yasin, S. Schafer, A. Arcones and A. Schwenk, Phys. Rev. Lett. {\bf 124}, 092701 (2020).
\bibitem{ozel16}
F. Ozel, P. Friere. Annu. Rev. Astron. Astrophys. {\bf 54}, 401 (2016).
\bibitem{bogd19}
S. Bogdanov, {\it et al.}, Astrophys. J. Lett. {\bf 887}, L25 (2019).
\bibitem{bogd19a}
S. Bogdanov, {\it et al.}, Astrophys. J. Lett. {\bf 887}, L26 (2019).
\bibitem{peik07}
J. Piekarewicz, Phys. Rev. C {\bf 76}, 031301 (R) (2007).
\bibitem{li07}
T. Li {\it et al.} Phys. Rev. L {\bf 99}, 162503 (2007).
\bibitem{cao12}
L.-G. Cao, H. Sagawa, G. Colò, Phys. Rev. C {\bf 86}, 054313 (2012).
\bibitem{khan07}
E. Khan, Phys. Rev. C {\bf 80}, 057302 (2009).
\bibitem{khan09}
E. Khan, Phys. Rev. C {\bf 80}, 011307 (2009).
\bibitem{howa20}
K. B. Howard {\it et al.} Phys. Lett. B {\bf 807}, 135608 (2020).
\bibitem{hara01}
M. N. Harakeh, A. van der Woude, {\it Giant Resonances: Fundamental High-Frequency Modes of Nuclear Excitation}, Oxford University Press, (New York, 2001).
\bibitem{garg18}
U. Garg, G. Colò, Prog. Part. Nucl. Phys. {\bf 101}, 55 (2018).
\bibitem{blai95}
J. P. Blaizot, J. F. Berger, J. Dechargé, M. Girod, Nucl. Phys. A {\bf 591} 435 (1995).
\bibitem{li08}
J. Li, G. Colò, J. Meng, Phys. Rev. C {\bf 78}, 064304 (2008).
\bibitem{shlomo06}
S. Shlomo, V. M. Kolomeitz and G. Colo, Eur. Phys. J. A {\bf 30}, 23 (2006).
\bibitem{li21}
B. A. Li and W. J. Xie, Phys. Rev. C {\bf 104}, 034610 (2021).
\bibitem{colo14}
G. Colò, U. Garg, and H. Sagawa, Eur. Phys. J. A {\bf 50}, 26 (2014).
 \bibitem{khan12}
E. Khan, J. Margueron, and I. Vidaña, Phys. Rev. Lett. {\bf 109}, 092501 (2012).
 \bibitem{marg18}
J. Margueron, R. Hoffmann Casali, and F. Gulminelli, Phys. Rev. C {\bf 97}, 025805 (2018).
\bibitem{dutra14}
M. Dutra, O. Lourenço, S. S. Avancini, B. V. Carlson, A. Delfino, D. P. Menezes, C. Providência, S. Typel, and J. R. Stone, Phys. Rev. C {\bf 90}, 055203 (2014).
\bibitem{patel12}
D. Patel, U. Garg, M. Fujiwara, H. Akimune, G. Berg, M. N. Harakeh, M. Itoh, T. Kawabata, K. Kawase, B. Nayak, T. Ohta, H. Ouchi, J. Piekarewicz, M. Uchida, H. Yoshida, M. Yosoi, Phys. Lett. B {\bf 718}, 447 (2012).
\bibitem{patel13}
D. Patel, U. Garg, M. Fujiwara, T. Adachi, H. Akimune, G. P. A. Berg, M. N. Harakeh, M. Itoh, C. Iwamoto, A. Long, J. T. Matta, T. Murakami, A. Okamoto, K. Sault, R. Talwar, M. Uchida, M. Yosoi, Phys. Lett. B {\bf 726}, 178 (2013).
\bibitem{jiang21}
C. L. Jiang, B. B. Back, K. E. Rehm, K. Hagino, G. Montagnoli and A. M. Stefanini, Eur. Phys. J. A {\bf 57}, 235 (2021).
\bibitem{back14}
B. B. Back, H. Esbensen, C. L. Jiang, and K. E. Rehm, Rev. Mod. Phys. {\bf 86}, 317 (2014).
\bibitem{gautam17}
M. S. Gautam, K. Vinod and H. Kumar, Braz. J. Phys. {\bf 47}, 461 (2017). 
\bibitem{pengo83}
R. Pengo, D. Evers, K. E. G. Lobner, U. Quade, K. Rudolph, S. J. SKorka and I. Weidl, Nucl. Phys. A {\bf 411}, 255 (1983).
\bibitem{tora17}
F. Torabi, O. N. Ghodsi and M. R. Pahlavani, Phys. Rev. C {\bf 95}, 034601 (2017).
\bibitem{gupta06}
R. K. Gupta, M. Manhas and W. Greiner, Phys. Rev. C {\bf 73}, 054307 (2006).
\bibitem{mont17}
G. Montagnoli and A. M. Stefanini, Eur. Phys. J. A {\bf 53}, 169 (2017).
\bibitem{toub17}
A. J. Toubiana, L. F. Canto and M. S. Hussein, Braz. J. Phys. {\bf 47}, 321 (2017).
\bibitem{das98}
M. Dasgupta, D. J. Hinde, N. Rowley and A. M. Stefanini, Annu. Rev. Nucl. Part. Sci. {\bf 48}, 401 (1998).
\bibitem{raj20}
Nuclear Structure Physics, edited by A. Shukla and S. K. Patra, (CRC Press, Boca Raton, 2020), Chapter 5.
\bibitem{canto20}
L. F. Canto, V. Guimaraes, J. Lubian and M. S. Hussein, Eur. Phys. J. A {\bf 56}, 281, (2020).
\bibitem{ghar22}
R. Gharaei and M. R. Yazdi, Nucl. Phys. A {\bf 1019}, 122381 (2022).
\bibitem{hagino98}
K. Hagino, N. Rowley, A.T.Kruppa, Comput. Phys. Commun. 123, 143 (1999).
\bibitem{deni20}
V.Yu. Denisov, Eur. Phys. J. A {\bf 7}, 87 (2000).
\bibitem{zagr03}
V. I. Zagrebaev, Phys. Rev. C {\bf 67}, 061601(R) (2003).
\bibitem{kohley13}
Z. Kohley, J. F. Liang, D. Shapira, C. J. Gross, R. L. Varner, J. M. Allmond, J. J. Kolata, P. E. Mueller, and A. Roberts
Phys. Rev. C {\bf 87}, 064612 (2013).
\bibitem{zhen19}
Z. Wu and L. Guo, Phys. Rev. C {\bf 100}, 024602 (2019).
\bibitem{jones22}
K. L. Jones, {\it et al.}, Phys. Rev. C {\bf 105}, 024602 (2022).
\bibitem{jiang15}
C. L. Jiang, {\it et al.}, Phys. Rev. C {\bf 91}, 044602 (2015).
\bibitem{liang07}
J. F. Liang{\it et al.}, Phys. Rev. C {\bf 75}, 054607 (2007).
\bibitem{misicu07}
Ş. Mişicu and H. Esbensen, Phys. Rev. C {\bf 75}, 034606 (2007).
\bibitem{misi06}
Ş. Mişicu and H. Esbensen, Phys. Rev. Lett. {\bf 96}, 112701 (2006).
\bibitem{esbe14}
H. Esbensen and A. M. Stefanini, Phys. Rev. C  {\bf 89}, 044616 (2014).
\bibitem{bhuy18}
M. Bhuyan and R. Kumar, Phys. Rev. C {\bf 98}, 054610 (2018).
\bibitem{rana22}
S. Rana, M. Bhuyan and R. Kumar, Phys. Rev. C {\bf 105}, 054613 (2022).
\bibitem{bhuy22}
M. Bhuyan, S. Rana, N. Jain, R. Kumar, S. K. Patra, and B. V. Carlson, Phys. Rev C {\bf 106}, 044602 (2022).
\bibitem{rana23}
S. Rana, M. Bhuyan, R. Kumar, and B. V. Carlson, Phys. Rev. C {\it communicated}, (2023).
\bibitem{meng16}
J. Meng, Relativistic Density Functional For Nuclear Structure, Word Scientific, Int. Rev. Nucl. Phys. {\bf 10}, (2016).
\bibitem{lala09}
G. A. Lalazissis, S. Karatzikos, R. Fossion, D. Pena Arteaga, A. V. Afanasjev and P. Ring, Phys. Lett. B {\bf 671}, 36 (2009).
\bibitem{ring96}
P. Ring, Prog. Part. Nucl. Phys. {\bf37}, 193 (1996).
\bibitem{biswal15}
S. K. Biswal, S. K. Singh and S. K. Patra, Mod. Phys. Lett. A {\bf 30}, 1550097 (2015).
\bibitem{biswal14}
S. K. Biswal and S. K. Patra, Cent. Eur. J. Phys. {\bf 12}, 582 (2014).
\bibitem{vret05}
D. Vretenar, A. V. Afanasjev, G. A. Lalazissis, P. Ring, Phys. Rep. {\bf 409}, 101 (2005).
\bibitem{kuma09}
R. Kumar, M. Bansal, S. Arun and R. K. Gupta, Phys. Rev. C {\bf 80}, 034618 (2009).
\bibitem{wong73}
C. Y. Wong, Phys. Rev. Lett. {\bf 31}, 766 (1973).
\bibitem{bock82}
R. Bock, Y. T. Chu, M. Dakowski et al., Nucl. Phys. A {\bf 388}, 334 (1982).
\bibitem{lesko86}
K. T. Lesko, W. Henning, K. E. Rehm, G. Rosner, J. P. Schiffer, G. S. F. Stephans, and B. Zeidman, Phys. Rev. C {\bf 34}, 2155 (1986).
\bibitem{raman03}
S. Raman, C. W. Nestor, JR. and P. Tikkanen, At. Data Nucl. Data Tables {\bf 78}, 1-128 (2001).
\bibitem{satc79}
G. R. Satchler and W. G. Love, Phys. Reports {\bf55}, 183 (1979).
\bibitem{sing12}
B. B. Singh, M. Bhuyan, S. K. Patra, and R. K. Gupta, J. Phys. G: Nucl. Part. Phys. {\bf39}, 025101 (2012).
\bibitem{sahu14}
B. B. Sahu, S. K. Singh, M. Bhuyan, S. K. Biswal, and S. K. Patra, Phys. Rev. C {\bf 84}, 034614 (2014).
\bibitem{piek09}
J. Piekarewicz, and M. Centelles, Phys. Rev. C {\bf 79}, 054311 (2009).
\bibitem{rutel05}
 B. G. Todd-Rutel and J. Piekarewicz, Phys. Rev. Lett. {\bf 95}, 122501 (2005).
 \bibitem{lala97}
  G. A. Lalazissis, J. Konig, and P. Ring, Phys. Rev. C {\bf 55}, 540 (1997).
\bibitem{rein86}
P. G. Reinhard, M. Rufa, J. Maruhan, W. Greiner and J. Friedrich, Z. Phys. A-At. Nucl. {\bf 323}, 15 (1986).
\bibitem{zeng83}
 J. Y. Zeng and T. S. Cheng, Nucl. Phys. A {\bf 405}, 1 (1983).
\bibitem{moli97}
 H. Molique and J. Dudek, Phys. Rev. C {\bf 56}, 1795 (1997).
\bibitem{zhang11}
 Z.-H. Zhang, J.-Y. Zeng, E.-G. Zhao, and S.-G. Zhou, Phys. Rev. C {\bf 83}, 011304(R) (2011).
\bibitem{hao12}
T. V. N. Hao, P. Quentin, and L. Bonneau, Phys. Rev. C {\bf 86}, 064307 (2012).
\bibitem{lala99}
 G. A. Lalazissis, D. Vretenar, P. Ring, M. Stoitsov, and L. M. Robledo, Phys. Rev. C {\bf 60}, 014310 (1999).
\bibitem{lala99a}
 G. A. Lalazissis, D. Vretenar, and P. Ring, Nucl. Phys. A {\bf 650}, 133 (1999).
\bibitem{doba84}
J. Dobaczewski, H. Flocard and J. Treiner, Nucl. Phys. A {\bf 422}, 103 (1984).
\bibitem{madl88}
D. G. Madland and J. R. Nix, Nucl. Phys. A {\bf 476}, 1 (1988).
\bibitem{patra11}
S. K. Patra, M. Del Estal, M. Centelles, and X. Vinas, Phys. Rev. C {\bf 63}, 024311 (2011).
\bibitem{moll16}
P. Möller, A. J. Sierk, T. Ichikawa , and H. Sagawa, At. Data Nucl. Data Tables {\bf109–110}, 1–204 (2016).
\bibitem{bohr52}
 A. Bohr, Mat. Fys. Medd. Dan. Vid. Selsk, {bf 26}, 14 (1952).
 \bibitem{bohr53}
A. Bohr and B. R. Mottelson, Mat. Fys. Medd. Dan. Vid. Selsk, {\bf 27}, 1 (1953). 
\bibitem{hill53}
D. L. Hill and J. A. Wheeler, Phys. Rev. {\bf 89}, 1102 (1953).
\bibitem{beck81} 
M. Beckerman, J. Ball, H. Enge, M. Salomaa, A. Sperduto, S. Gazes, A. DiRienzo, and J. D. Molitoris, Phys. Rev. C {\bf 23}, 1581 (1981).
\bibitem{long08}
Z. Gao-Long and L. Xiao-Yun, Chin. Phys. C {\bf 32}, 812 (2008).
\bibitem{lari16}
 F. Lari and O.N. Ghodsi,  Commun. Theor. Phys. {\bf 65}, 213 (2016).
\bibitem{rash96}
M. Rashdan, J. Phys. G: Nucl. Part. Phys. {\bf 22}, 139 (1996).
\bibitem{arit12}
Y. Aritomo, K. Hagino, K. Nishio and S. Chiba, Phys. Rev. C {\bf 85}, 044614 (2012).
\bibitem{sun23}
X. Sun and L.  Guo, Phys. Rev. C {\bf 107}, L011601 (2023).
\bibitem{rein02}
P.-G. Reinhard, M. Bender and J. A. Maruhn, nucl-th/0012095.
\bibitem{afan05}
A. V. Afanasjev and S. Frauendrof, Phys. Rev. C {\bf 71}, 024308 (2005).
\bibitem{chu10}
Y. Chu, Z. Ren, Z. Wang and T. Dong, Phys. Rev. C {\bf 82}, 024320 (2010).
\bibitem{gupta07}
R. K. Gupta, D. Singh and W. Greiner, Phys. Rev. C {\bf 75}, 024603 (2007).
\bibitem{ghodsi16}
O. N. Ghodsi and F. Torabi, Phys. Rev. C {\bf 93}, 064611 (2016).
\bibitem{vijay22}
Vijay, N. Grover, K. Sharma, M. S. Gautam, M. K. Sharma, and R. P. Chahal, Phys. Rev. C {\bf 106}, 064609 (2022).
\end{thebibliography}
\end{document}